# Mechanical model of the ultra-fast underwater trap of *Utricularia*

Marc Joyeux[(#)], Olivier Vincent and Philippe Marmottant[(¶)]

*Laboratoire de Spectrométrie Physique (CNRS UMR5588),*

*Université Joseph Fourier Grenoble 1, BP 87, 38402 St Martin d'Hères, France*

**Abstract** : The underwater traps of the carnivorous plants of the *Utricularia* species catch their preys through the repetition of an "active slow deflation / passive fast suction" sequence. In this paper, we propose a mechanical model that describes both phases and strongly supports the hypothesis that the trap door acts as a flexible valve that buckles under the combined effects of pressure forces and the mechanical stimulation of trigger hairs, and not as a panel articulated on hinges. This model combines two different approaches, namely (i) the description of thin membranes as triangle meshes with strain and curvature energy, and (ii) the molecular dynamics approach, which consists in computing the time evolution of the position of each vertex of the mesh according to Langevin equations. The only free parameter in the expression of the elastic energy is the Young's modulus $E$ of the membranes. The values for this parameter are unequivocally obtained by requiring that the trap model fires, like real traps, when the pressure difference between the outside and the inside of the trap reaches about 15 kPa. Among other results, our simulations show that, for a pressure difference slightly larger than the critical one, the door buckles, slides on the threshold and finally swings wide open, in excellent agreement with the sequence observed in high-speed videos.

PACS numbers :

87.19.R-         : mechanical and electrical properties of tissues and organs

87.10.Tf         : biological and medical physics / molecular dynamics simulations

62.20.mq       : mechanical properties of solids / buckling.

[(#)] email : Marc.Joyeux@ujf-grenoble.fr

[(¶)] email : Philippe.Marmottant@ujf-grenoble.fr

# 1 - Introduction

There exist more than 600 species of carnivorous plants, which are the result of adaptation to poor environments in terms of nutriments and/or sunshine [1]. The various methods used by these plants to catch animals may be divided into two main categories, namely active and passive traps, depending on whether capture of a prey does or does not involve any motion of the plant itself. Plants of the genus Nepenthes are typical examples of carnivorous plants with passive traps. Their catching mechanism relies mostly on the shape of the pitcher-like sleeves and the high viscoelasticity of the digestive fluid [2]. In contrast, the closure of the Venus flytrap leaf in about 100 ms following mechanical stimulation of trigger hairs is a well-known example of active trap [3].

Of the various active traps, none has, however, intrigued botanists more than those of the about 215 species of *Utricularia* [4-14]. These traps are aquatic, millimeter-sized, lenticular bladder-like organs [15,16] (see fig. 1a). They have an entrance, which remains closed by a door most of the time (see fig. 1b-1c). Firing of the *Utricularia* trap is a two-steps mechanism. During the first, slow step, the door is indeed closed and particular glands actively pump water out of the trap interior. This has two consequences. First, the hydrostatic pressure inside the trap drops below that outside the trap by about 10-20 kPa [11,13]. Moreover, the concave wall curvature due to the lower internal pressure results in elastic energy being stored in the walls. We will show in the next section that this deflation step is an essentially exponential process with a time constant of about 1 hour. The second, ultra-fast step starts when a potential prey touches one of the trigger hairs attached close to the center of the door. Then the door opens, and water (and the prey) are engulfed while the walls of the trap release the stored energy and relax to their equilibrium position. When the pressures inside and outside the trap are levelled, the door closes again autonomously.

We recently used a combination of high-speed video imaging, scanning electron microscopy, light-sheet fluorescence microscopy, particle tracking, and molecular dynamics simulations, to visualize the motion of the door and propose a plausible mechanism. In particular, we observed that the time span of suction is smaller than one millisecond, that is, substantially shorter than previously estimated [4]. We also measured a maximum liquid velocity of about 1.5 m s$^{-1}$ and a maximum acceleration of 600 g, which leave little escape chances to small preys. More importantly, our high-speed video recordings (up to about 10000 frames per second), in combination with light-sheet microscopy, reveal that the opening of the door is preceded by the inversion of its curvature, and not the opposite as was previously assumed [8]. After excitation of the trigger hairs, the (initially convex) door indeed bulges inside and becomes concave, starting at the area of trigger hair insertion. It is only when this inversion of curvature has spread over the whole door surface that the door opens and swings inside very rapidly. These videos, which will be made available as Supplementary Material of a separate article [17], therefore suggest that the extremely fast opening of the door is similar to the buckling of a flexible valve [18] rather than the rotation of an almost rigid panel articulated on hinges.

Most part of these experimental results will be published in a biology journal [17]. The purpose of the present complementary article is to show that the hypothesis of door buckling is confirmed by molecular dynamics simulations based on the description of the body and the door of the trap as thin membranes with Young's moduli in the range 2-10 MPa. We will provide a complete description of the model and the results obtained there with.

We actually first show in sect. 2 that important information can be extracted from experimental results by using a very simple model, which consists of two parallel disks connected by a spring. The remainder of the paper is then devoted to the description of the membrane model and the discussion of the results obtained there with. For the sake of faster calculations, we separated simulations concerning the body of the trap from those concerning the door and developed two different models, which however share many ingredients. The ingredients that are common to both models, that is, the expressions of the potential energy of the membrane and the equations of evolution, are presented in sect. 3. The model for the trap body is then discussed in sect. 4, and that for the door in sect. 5.

**2 - A first approach : the disks-and-spring model**

In this section, we show that a very simple, scalar model enables to extract important information from experimental data.

2.1 - *Setting of the trap (deflation phase)*

The model consists in describing the body of a trap as two parallel disks of diameter *L* separated by a distance *e* and connected by a spring of constant *k*. It is assumed that the geometry of the trap remains that of a cylinder, that is, an impermeable and highly extensible membrane closes the volume between the two disks. *e* represents the thickness of the trap. Experimentally, the traps are viewed from above (that is, along the *x* axis of Fig. 1) and their thickness *e* is measured close to the center of the body, as is shown in the inset in the top plot of fig. 2. A typical curve for the time evolution of *e* during setting (deflation) of an *Utricularia inflata* trap is shown in fig. 2 on linear (top plot) and logarithmic (bottom plot) scales. The trap is fired manually and measurement of *e* starts immediately after the ultra-fast opening and closing of the door. The thickness of the trap is therefore maximum at $t = 0$. Figure 2 indicates that *e* evolves exponentially with time, according to

$$e(t) = e_{min} + (e_{max} - e_{min})\exp(-\frac{t}{\tau_{pump}}) \; , \qquad (2.1)$$

where $e_{max}$ is the thickness of the trap at rest (completely inflated), $e_{min}$ its thickness when it is completely deflated and ready to fire, and $\tau_{pump}$ is the characteristic time for pumping. For the trap and the deflation event shown in fig. 2, we measured $e_{max} \approx 0.80$ mm, $e_{min} \approx 0.37$ mm, and $\tau_{pump} \approx 53$ minutes. Successive experiments performed with this same trap led to values of $\tau_{pump}$ that varied by less than the uncertainty of the fit, that is a few minutes. In contrast, measurements performed with different traps led to rather different values of $\tau_{pump}$, which ranged from 28 to 53 minutes. This large scattering in the values of $\tau_{pump}$ is certainly due, in part, to differences in the size of the investigated traps, but it may also result from different efficiencies of the respective sets of pumping glands. We also note in passing that the large value of the characteristic time for pumping obliged us to wait several hours between two successive experiments performed on the same trap, in order for the trap to be always in the same (almost) steady state when fired.

The maximum pumping rate $Q_0$, that is, the pumping rate at $t=0$, can furthermore be estimated from

$$Q_0 = -\left(\frac{dV}{dt}\right)_{t=0} = \frac{\pi}{4} L^2 \frac{e_{max} - e_{min}}{\tau_{pump}}, \tag{2.2}$$

where $V$ is the volume comprised between the two disks. When plugging in eq. (2.2) the value $L = 1.5$ mm, as well as those derived above for $e_{min}$, $e_{max}$, and $\tau_{pump}$, one obtains $Q_0 \approx 0.86$ mm$^3$ hr$^{-1}$, which compares well with the value reported in ref. [13], that is, $Q_0 \approx 1.26$ mm$^3$ hr$^{-1}$. An upper limit for the hydraulic permeability of the trap walls, $\kappa_h$, can furthermore be estimated by assuming that the pumping rate is constant and equal to $Q_0$, and that transfers of liquid between the inside and the outside of the trap arise uniquely from the porosity of the walls. Then

$$\kappa_h = \frac{\eta Q_0 h}{2S \Delta p_{max}} = \frac{2\eta Q_0 h}{\pi L^2 \Delta p_{max}} \;, \qquad (2.3)$$

where $\eta$ is the viscosity of the fluid ($\eta \approx 10^{-3}$ Pa s), $h$ the thickness of the wall, $S$ the surface of each disk, and $\Delta p_{max}$ the steady-state pressure difference between the inside and the outside of the trap. When plugging $h \approx 100$ μm and $\Delta p_{max} \approx 15$ kPa [11,13] in eq. (2.3), one gets $\kappa_h \approx 45$ Å$^2$. At last, the constant $k$ of the spring is such that pressure forces $2S \Delta p$ and the spring elastic force $k(e_{max} - e_{min})$ cancel at maximum deflation, that is when $\Delta p = \Delta p_{max}$ and $e = e_{min}$. One therefore has

$$k = \frac{2S \Delta p}{e_{max} - e_{min}} = \frac{\pi L^2 \Delta p}{2(e_{max} - e_{min})} \;, \qquad (2.4)$$

which leads to $k = 120$ J m$^{-2}$. The elastic energy stored in the membrane during the deflation phase, $\frac{k}{2}(e_{max} - e_{min})^2$, is consequently close to 11 μJ.

### 2.2 - *Firing of the trap (inflation phase)*

Let us now consider the inflation of the trap once it is manually triggered and the door opens. A typical curve for the time evolution of $e$ during the suction phase (inflation) of an *Utricularia inflata* trap is shown in fig. 3 on linear (top plot) and logarithmic (bottom plot) scales. Figure 3 indicates that the time evolution of $e$ is not mono-exponential, but most of the gap to maximum thickness (or volume) is nevertheless bridged with a time constant of the order of 1 ms. Moreover, the maximum speed of the walls of the trap can be estimated by taking the numerical derivative of the curve in the top plot of fig. 3. One obtains $(de/dt)_{max} \approx 0.14$ m s$^{-1}$. The variation of $e$ can be related to the average speed $u$ of the fluid entering the trap by considering that the door is a disk of radius $r = 300$ μm. Conservation of volume then implies that

$$\frac{dV}{dt} = \pi r^2 u , \qquad (2.5)$$

which can be rewritten in the form

$$u = \left(\frac{L}{2r}\right)^2 \frac{de}{dt} . \qquad (2.6)$$

The maximum value of $u$ deduced from the plots in fig. 3 is therefore $u_{max} \approx 0.9$ m s$^{-1}$. One thereby recovers in a comparatively simpler way the result obtained by tracking the motion of hollow glass beads of density 1.1 and diameter 6-20 µm initially dispersed in the fluid. The motion of these tracers during the suction phase was recorded using a high-speed Phantom Miro 4 camera (up to 8100 frames per second for images with 256*256 pixels) placed on the side of the traps, that is along the $z$ axis. These more elaborate experiments lead to a maximum speed of the fluid of about 1.5 m s$^{-1}$ [17]. They additionally indicate that the acceleration of the fluid reaches the impressive value of 600g.

The maximum Reynolds number along the flow, Re, writes

$$\mathrm{Re} = \frac{2 r u_{max}}{\nu} , \qquad (2.7)$$

where $\nu = 10^{-6}$ m$^2$ s$^{-1}$ is the kinematic viscosity of water. One obtains $\mathrm{Re} \approx 540$, which indicates that the flow entering the trap is strongly inertial but still remains laminar, since fully developed turbulence arises only for Reynolds numbers larger than 2000 [19].

At last, one may estimate the characteristic inertial time for trap inflation, $\tau_i$, by considering that it is equal to one half of the oscillation period of a mass $m$ (equal to the mass of one disk) attached to a spring with the constant $k$ determined above, that is,

$$\tau_i = \frac{\pi}{2} \sqrt{\frac{m}{k}} . \qquad (2.8)$$

The inertia of an object is larger in a liquid than in air, because of the mass of the liquid that is displaced during the motion of the object. Therefore, $m$ can be estimated as the mass of the liquid that is displaced by each disk during inflation and deflation, that is, $m = \frac{1}{2}\rho S(e_{max} - e_{min})$, where $\rho$ is the density of water. One obtains $m \approx 0.4$ mg, and consequently $\tau_i \approx 0.2$ ms. This estimate of $\tau_i$ is one order of magnitude smaller than the time it actually takes for the trap to inflate (see fig. 3). This indicates that friction plays a crucial dynamical role in slowing down the inflation motion from the 0.1 ms time scale to the 1 ms one. We will come back later to this point.

The very simple disks-and-spring model therefore enables one to estimate some of the principal characteristics of the trap, namely the maximum pumping rate (about 1 mm$^3$ hr$^{-1}$), the characteristic pumping time (about 1 hr), the hydraulic permeability of the membrane (a few tens of Å$^2$), and the average elastic energy stored in the membrane (in the µJ range). Moreover, it leads to the correct value for the maximum velocity of the fluid (about 1 m s$^{-1}$), and suggests that the observed time scale of the dynamics (a few ms) is imposed by the frictions with the surrounding liquid and not by the inertia of the trap body itself. However, this model provides no indication concerning the actual mechanisms that enable such astounding catching performances. This is essentially due to the fact that it describes the body of the trap but completely disregards the door, which is certainly the most intriguing part of this plant. We therefore developed a more elaborate membrane model, in order to get a better understanding of the dynamics of the trap.

### 3 - Membrane model

The remainder of this article is devoted to the description of the 3-dimensional membrane model and the discussion of the results obtained there with. For the sake of faster calculations, we separated simulations concerning the body of the trap from those concerning its door and developed two different models, which however share many ingredients. We describe in the present section the ingredients that are common to both models, that is, the expressions of the potential energy and the equations of evolution, as well as the discretization procedure. We postpone the complete presentation of the model for the trap body to sect. 4, and that for the door to sect. 5.

Both the trap body and the door are modelled as thin membranes of thickness $h$, which are made of an isotropic, homogeneous, and incompressible material with Young's modulus $E$ and Poisson ratio $\nu = \frac{1}{2}$. Note, however, that the Young's moduli of the body and the trap are not necessarily identical, because they are made of cells with different thickness and different spatial organisation. The elastic potential energy stored in the deformation of the membrane, $E_{pot}$, can be written as the sum of a strain contribution, $E_{strain}$, and a curvature contribution, $E_{curv}$, according to [20,21]

$$E_{pot} = E_{strain} + E_{curv}$$
$$E_{strain} = \frac{E h}{2(1-\nu^2)} \int_S [(1-\nu)\text{Tr}(\boldsymbol{\varepsilon}^2) + \nu(\text{Tr}(\boldsymbol{\varepsilon}))^2] \, dS \qquad (3.1)$$
$$E_{curv} = \frac{E h^3}{24(1-\nu^2)} \int_S [(\text{Tr}(\mathbf{b}))^2 - 2(1-\nu)\text{Det}(\mathbf{b})] \, dS \,,$$

where $S$ is the area of the membrane, $\boldsymbol{\varepsilon}$ the 2-dimensional Cauchy-Green local strain tensor [22], and $\mathbf{b}$ the difference between the local curvature tensors of the strained membrane and the reference geometry (see below). For numerical purposes, all membranes are described as triangle meshes with $M$ triangles (facets) and $N \approx M/2$ vertices. Denoting by $\delta S_n$ the area of facet $n$, the elementary area $\delta A_j$ associated to vertex $j$ is

$$\delta A_j = \frac{1}{3} \sum_{n \in V_1(j)} \delta S_n \,, \qquad (3.2)$$

where $n \in V_1(j)$ means that the sum runs over all the facets $n$ that contain vertex $j$. Each vertex $j$ is also associated a mass $m_j$, which is derived from the reference geometry according to

$$m_j = \rho h \delta A_j, \tag{3.3}$$

where $\rho$ is the density of the membrane. We used $\rho = 1$ kg dm$^{-3}$, because the cells that form the membrane are filled with water and the trap itself is very close to the floating limit. Use of a different value for $\rho$ would only modify the kinetic energy proportionally and would not change qualitatively the results presented below. The mass $m_j$ of each vertex is then kept constant during the simulations, while area elements $\delta S_n$ and $\delta A_j$ may vary.

$E_{\text{strain}}$ is discretized in the form

$$E_{\text{strain}} = \frac{Eh}{2(1-\nu^2)} \sum_{n=1}^{M} [(1-\nu)\text{Tr}(\varepsilon_n^2) + \nu(\text{Tr}(\varepsilon_n))^2] \delta S_n, \tag{3.4}$$

where the Cauchy-Green strain tensor [22] for facet $n$, $\varepsilon_n$, writes

$$\varepsilon_n = \frac{1}{2}(\mathbf{F}_n \cdot (\mathbf{F}_n^0)^{-1} - \mathbf{I}). \tag{3.5}$$

In this equation, $\mathbf{I}$ denotes the 2×2 identity matrix, while $\mathbf{F}_n$ and $\mathbf{F}_n^0$ are the Gram matrices for facet $n$ in the strained geometry and the reference one, that is,

$$\mathbf{F}_n = \begin{pmatrix} (\mathbf{r}_{n2} - \mathbf{r}_{n1}) \cdot (\mathbf{r}_{n2} - \mathbf{r}_{n1}) & (\mathbf{r}_{n2} - \mathbf{r}_{n1}) \cdot (\mathbf{r}_{n3} - \mathbf{r}_{n1}) \\ (\mathbf{r}_{n2} - \mathbf{r}_{n1}) \cdot (\mathbf{r}_{n3} - \mathbf{r}_{n1}) & (\mathbf{r}_{n3} - \mathbf{r}_{n1}) \cdot (\mathbf{r}_{n3} - \mathbf{r}_{n1}) \end{pmatrix}, \tag{3.6}$$

where $\mathbf{r}_{n1}$, $\mathbf{r}_{n2}$, and $\mathbf{r}_{n3}$ describe the positions of the three vertices of the facet.

The contribution to energy arising from curvature, $E_{\text{curv}}$, is more difficult to evaluate. The terms containing $\text{Tr}(\mathbf{b})$ and $\text{Det}(\mathbf{b})$ in eq. (3.1) are known as the mean curvature energy and the Gaussian curvature energy, respectively. They can be rewritten in the more explicit form

$$\begin{aligned} E_{\text{curv}} &= E_{\text{mean}} + E_{\text{Gauss}} \\ E_{\text{mean}} &= \frac{Eh^3}{24(1-\nu^2)} \int_S (c_1 + c_2 - c_1^0 - c_2^0)^2 \, dS \\ E_{\text{Gauss}} &= -\frac{Eh^3}{12(1+\nu)} \int_S ((c_1 - c_1^0)(c_2 - c_2^0) - \sin^2\theta (c_1^0 - c_2^0)(c_1 - c_2)) \, dS, \end{aligned} \tag{3.7}$$

where the $c_k$ and $c_k^0$ ($k=1,2$) are the local principal curvatures of the strained membrane and those of the reference geometry, respectively, and $\theta$ is the angle by which the local principal directions of the membrane have rotated with respect to those of the reference geometry. The mean curvature energy is rather straightforwardly discretized according to

$$E_{\text{mean}} = \frac{Eh^3}{6(1-\nu^2)} \sum_{j=1}^{N} (\kappa_j - \kappa_j^0)^2 \delta A_j \ . \tag{3.8}$$

In this equation, $\kappa_j$ and $\kappa_j^0$ represent the mean curvature $\kappa = (c_1 + c_2)/2$ at vertex $j$ for the strained membrane and the reference geometry, respectively. They are estimated from [23,24]

$$\kappa_j = \frac{1}{4\delta A_j} \left\| \sum_{k \in N_1(j)} (\cot \alpha_{jk} + \cot \beta_{jk})(\mathbf{r}_k - \mathbf{r}_j) \right\| \ , \tag{3.9}$$

where $k \in N_1(j)$ means that the sum runs over the vertices $k$ that are directly connected to vertex $j$. $\mathbf{r}_j$ and $\mathbf{r}_k$ denote the positions of vertices $j$ and $k$, and $\alpha_{jk}$ and $\beta_{jk}$ are the angles of the corners opposite to bond ($jk$) in the two facets that share this bond. The problem actually arises from the Gaussian curvature energy, because it is difficult to estimate $\theta$ correctly in the course of a simulation. We consequently used an approximate expression for $E_{\text{Gauss}}$, namely

$$E_{\text{Gauss}} \approx -\frac{Eh^3}{12(1+\nu)} \int_S (c_1 c_2 - c_1^0 c_2^0 - \frac{1}{2}(c_1^0 + c_2^0)(c_1 + c_2 - c_1^0 - c_2^0)) dS \ . \tag{3.10}$$

Note that it is sufficient that $c_1^0 - c_2^0$ be equal to zero everywhere on the membrane for the expressions for $E_{\text{Gauss}}$ in eqs. (3.7) and (3.10) to be equivalent. This is the case, in particular, if the membrane has no spontaneous curvature ($c_1^0 = c_2^0 = 0$ everywhere) or if the reference geometry is a sphere of radius $R$ ($c_1^0 = c_2^0 = 1/R$ everywhere). Equation (3.10) is finally discretized according to

$$E_{\text{Gauss}} \approx -\frac{Eh^3}{12(1+\nu)} \sum_{j=1}^{N} (G_j - G_j^0 - 2\kappa_j^0(\kappa_j - \kappa_j^0)) \delta A_j \ . \tag{3.11}$$

In eq. (3.11), $G_j$ and $G_j^0$ represent the Gaussian curvature $G = c_1 c_2$ at vertex $j$ for the strained membrane and the reference geometry, respectively, which we estimate from [25]

$$G_j = \frac{1}{\delta A_j}(2\pi - \sum_{n \in V_1(j)} \gamma_{nj}) , \qquad (3.12)$$

where $\gamma_{nj}$ denotes the angle at vertex $j$ in facet $n$.

At that point, the important question that arises is: what are the reference geometries, that is, those for which the Gram matrices $\mathbf{F}_n^0$ and spontaneous curvatures $\kappa_j^0$ and $G_j^0$ must be calculated ? In order to answer this question, we cut several sections of the trap body and the door and observed the resulting shapes. Two examples are shown in fig. 4. Figure 4a shows a transverse section of the trap body, while fig. 4b shows the door, which has been separated from the rest of the trap, seen from the edge that rests on the threshold. Conclusion of these experiments is that these parcels certainly do not become flat, but retain instead essentially the shape of the inflated trap. Stated in other words, the $\mathbf{F}_n^0$, $\kappa_j^0$, and $G_j^0$ must be computed for a geometry which is close to the equilibrium one when the pressure outside the trap is equal to that inside. The fact that the spontaneous curvatures are different from 0 has two important consequences. At first, this implies that the Gaussian curvature energy does not reduce to the integral of $c_1 c_2$, so that it does not remain constant upon deformation, even in the case of a closed surface (note, however, that for the closed surface describing the trap body, the Gauss-Bonnet theorem insures that the sum over $j$ of $G_j - G_j^0$ in eq. (3.11) remains constant upon deformation). Moreover, when estimating the Gaussian curvature energy according to eq. (3.10), the potential energy is not necessarily exactly minimum for the reference geometry, for which the $\mathbf{F}_n^0$ and $\kappa_j^0$ and $G_j^0$ are calculated. Once these quantities have been calculated, the geometry with minimum potential energy, which corresponds to the system at rest, must therefore be searched for. It usually differs only slightly from the reference geometry.

A proper investigation of the dynamics of the *Utricularia* trap would require the consideration of explicit liquid in addition to the membrane discussed above. Motion of the fluid would be described by Navier-Stokes equations and that of the membrane by Hamilton or Newton equations. The motion of the membrane and that of the liquid would be coupled through the pressure forces and the frictions exerted by the liquid on the membrane. This is, however, a very complex problem. We actually chose a simpler approach, which consists in solving Langevin equations for the membrane. More precisely, the position $\mathbf{r}_j$ of each vertex $j$ is assumed to satisfy

$$m_j \frac{d^2 \mathbf{r}_j}{dt^2} = -\nabla E_{\text{pot}} - \Delta p \, \delta A_j \mathbf{n}_j - m_j \gamma \frac{d\mathbf{r}_j}{dt} + \sqrt{2 m_j \gamma k_B T} \frac{dW(t)}{dt} \ . \tag{3.13}$$

In this equation, $\Delta p$ is the pressure outside the trap minus the pressure inside, $\mathbf{n}_j$ the outward normal to the surface at vertex $j$, $\gamma$ the dissipation coefficient, and $W(t)$ a Wiener process. The first and second term in the right-hand side of eq. (3.13) describe elastic and pressure forces, respectively, while the two last terms model the effects of the liquid, namely friction and thermal noise. Note that thermal noise (the last term) is negligibly small compared to elastic and pressure forces. $\mathbf{n}_j$ is computed according to

$$\mathbf{n}_j = \frac{\sum_{n \in V_1(j)} \delta S_n \mathbf{u}_n}{\left\| \sum_{n \in V_1(j)} \delta S_n \mathbf{u}_n \right\|} \ , \tag{3.14}$$

where $\mathbf{u}_n$ is the outward normal to facet $n$. For numerical purposes, the derivatives in Langevin equations are replaced by finite differences. The position of vertex $j$ at time step $i+1$, $\mathbf{r}_j^{i+1}$, is consequently obtained from the positions $\mathbf{r}_j^i$ and $\mathbf{r}_j^{i-1}$ at the two previous time steps according to

$$m_j(1+\frac{\gamma\Delta t}{2})\mathbf{r}_j^{i+1} = 2m_j\mathbf{r}_j^i - m_j(1-\frac{\gamma\Delta t}{2})\mathbf{r}_j^{i-1} - (\nabla E_{\text{pot}} + \Delta p\,\delta\!A_j\mathbf{n}_j)\Delta t^2 + \sqrt{2m_j\gamma k_B T}\,\Delta t^{3/2}w(t),$$

(3.15)

where $\Delta t$ is the time step and $w(t)$ a normally distributed random function with zero mean and unit variance.

It is important to realize that the model described above actually depends on two adjustable parameters, namely the Young's modulus $E$, which determines the strength of the elastic energy of the membrane in eq. (3.1), and the dissipation coefficient $\gamma$, which determines the strength of liquid/membrane interactions in Langevin equations (3.13). On the other side, experiments yield two fundamental quantities, namely the pressure difference $\Delta p$ in set conditions ($\Delta p$ is in the range 10-20 kPa [11,13]), and the time scales at which the door opens (a few tenths of ms) and the trap inflates (a few ms). As will be shown below, the Young's moduli of the trap and door membranes can be unambiguously derived from the experimental value of $\Delta p$, while the value of $\gamma$ is obtained by requiring that the door opening and trap inflation time scales computed with the model match the observed ones. This is therefore a very favorable case, where all the parameters of the model can be deduced from experiment.

**4 - Dynamics of the trap body**

As already mentioned, we separated, for the sake of faster calculations, simulations performed for the trap body from those concerning the door. In this section, we describe the model we developed for the trap body and the results obtained there with. The model for the door will be discussed in the following section.

### 4.1 - *Geometry of the trap*

The trap body is modelled as a closed shell of thickness $h = 100$ µm. It contains no aperture. The setting phase (deflation) is simulated by decreasing slowly the internal pressure relative to the external one. Once the trap is set, firing and the subsequent inflation are simulated by resetting instantly to zero the pressure difference between the inside and the outside of the trap.

The first question that arises is that of the geometry of the trap body. By considering the shape of real traps, like the one shown in fig. 1a, we first described the inflated trap as an oblate ellipsoid with major radius of 1 mm and minor radius in the range 0.5-0.7 mm. However, results obtained with this geometry differ markedly from the observed behavior, for all realistic values of the Young's modulus $E$ and the dissipation coefficient $\gamma$. Such simulations indeed predict that deflation consists of a single, abrupt buckling of the membrane, while observation instead leads to the conclusion that deflation is an essentially smooth and continuous process, although it seems that some limited buckling of small portions of the surface sometimes occur. This difference is due to the fact that the real trap is not convex everywhere but contains instead regions with negative curvature even in the inflated geometry. These regions with negative curvature actually act as seeds from which deflation propagates like a rolling wave when the internal pressure is decreased.

Therefore, we introduced such regions with negative curvature in our model by considering that the geometry of the inflated trap is obtained by transforming the coordinates $(x_j, y_j, z_j)$ of the vertices of a triangulated sphere of radius 1 mm according to

$$\begin{pmatrix} x_j \\ y_j \\ z_j \end{pmatrix} \to \begin{pmatrix} x_j \\ y_j - 0.1(y_j^2 - z_j^2) \\ 0.55\, z_j - 0.12\,(1 - x_j^2)(\sin(\frac{3}{2})z_j + 2\sin(3)y_j z_j + \sin(\frac{9}{2})z_j(3y_j^2 - z_j^2) + 4\sin(6)y_j z_j(y_j^2 - z_j^2)) \end{pmatrix}$$

(4.1)

The transformation of eq. (4.1) may look rather arbitrary, especially for the $z_j$ coordinate. However, the trigonometric terms that appear in this equation are just the first terms of the Fourier expansion of a Dirac peak and are aimed at creating a region with negative curvature on both sides of the trap. Moreover, the overall shape of the trap body is convincingly reproduced with this expression. The $\mathbf{F}_n^0$, $\kappa_j^0$, and $G_j^0$ are calculated for this geometry and the geometry corresponding to the minimum of the potential energy is then searched for. From the practical point of view, and unless otherwise stated, we used a mesh with about $N \approx 2150$ vertices and $M \approx 4300$ facets. For this mesh, the minimum energy geometry corresponds to a potential energy $E_{\text{pot}} \approx -0.89$ µJ and is only slightly different from that described by eq. (4.1). It is shown in the left picture of fig. 5. The area with negative curvature is clearly seen on the side of the body (the surface being symmetric with respect to the $xy$ plane, there obviously exists a similar area with negative curvature on the hidden side). If the model would contain a door, then this door would face the right edge of the picture, as in fig. 1a.

4.2 - *Setting of the trap (deflation phase)*

We then determined the value of the Young's modulus $E$ by requiring that maximum deflation, corresponding to a reduced volume $V_{\text{red}} \approx 0.6$, is achieved for a pressure difference $\Delta p \approx 15$ kPa [11,13] (the reduced volume $V_{\text{red}}$ is defined as the actual volume of the strained trap divided by the volume of the minimum energy geometry). To this end, we decreased the pressure inside the trap at the "slow" rate of 1 Pa µs$^{-1}$ and integrated Langevin equations (3.15) for $\gamma = 0$ with a time step $\Delta t = 10$ ns.

We obtained that the Young's modulus of the trap membrane is about $E = 7.2$ MPa, which lies in the range of values that are commonly measured for parenchymatous tissues (see for example refs. [26-28]). The bottom plot of fig. 6 shows the evolution of $\Delta p$ as a function of $1-V_{red}$. It can be checked on this plot that $V_{red}$ is indeed close to 0.6 for $\Delta p \approx 15$ kPa. The geometry of the trap for $V_{red}=0.6$, that is, in set-conditions, is displayed in the right picture of fig. 5. In addition, the top plot of fig. 6 shows the evolution of the elastic energy stored in the membrane, $E_{pot}$, as a function of $1-V_{red}$. It is seen that the available elastic energy in set-conditions is of the order of several µJ, in fair agreement with the estimation obtained from the disks-and-spring model.

The deflation curve obtained with $\gamma=0$ looks smooth and continuous (see fig. 6). This is, however, no longer the case for the deflation curve obtained with $\gamma=10^4$ s$^{-1}$, which is also shown in fig. 6. For this value of the dissipation coefficient, the deflation curve displays several plateaus, which are the fingerprints of a series of small bucklings, which involve limited portions of the body membrane. This indicates that the minimum energy pathway that leads from the inflated to the deflated trap actually consists of several (many ?) minima separated by energy barriers. When $\gamma=0$, the system acquires sufficient kinetic energy to surf above these barriers. When $\gamma>0$, the kinetic energy, which is released each time the system overcomes a barrier leading to a deeper minimum, is instead dissipated and the system may remain blocked in this minimum till sufficient pressure work has again been brought to him. Also drawn in the top plot of fig. 6 is the "static" curve obtained by minimizing $E_{pot}$ (with the conjugated gradient method) for increasing values of $V_{red}$. It can be observed that the "dynamic" curves obtained by integrating Langevin equations are always located slightly but significantly above the static one, which confirms that the actual pathway for deflation does not coincide exactly with the minimum energy pathway.

At that point, it is worth emphasizing that the precise sequence of buckling events depends on the mesh. In particular, the finer the mesh, the smaller the amplitude of buckling events, and the more continuous the deflation process. This is clearly seen in the bottom plot of fig. 6, which displays curves obtained with $\gamma = 10^4$ s$^{-1}$ for two different meshes, namely the standard one with about 2150 vertices and a rougher one with only about 200 vertices. For the rougher mesh, the number of plateaus is approximately divided by two compared to the finer one, but these plateaus are wider and the steps are higher. Since small bucklings are also observed during the deflation phase of real *Utricularia* traps, this suggests that the cells that form the membrane play approximately the same role as the mesh in our simulations, and that their rather large size is actually responsible for the observed buckling events.

### 4.3 - *Firing of the trap (inflation phase)*

Let us now turn our attention to the inflation phase, that is, the firing of the trap. After the trigger hairs have been excited, the door opens completely in about 0.5 ms. Due to the combined actions of pressure forces and the relaxation of the walls of the trap to their equilibrium positions, thereby releasing the stored elastic energy, water (and the eventual prey) are engulfed. Once the pressures inside and outside the trap are levelled, the door closes again autonomously. The whole process lasts a few milliseconds (see fig. 3). In this section, this is simply modelled by assuming that the trap is initially at equilibrium with a pressure difference $\Delta p \approx 15$ kPa and a reduced volume $V_{\text{red}} = 0.6$, that is, in the configuration shown in the right picture of fig. 5, and that at time $t = 0$ the pressure difference is instantly set to $\Delta p = 0$. Langevin equations (3.15) are then integrated with a time step $\Delta t = 2.5$ ns.

The time evolution of $V_{\text{red}}$ obtained from a simulation with a dissipation coefficient $\gamma = 0$ is shown in the top plot of fig. 7. This simulation agrees qualitatively with the disks-and-spring model described in section 2, in the sense that it predicts that the characteristic period of the free motion of the trap is of the order of 0.2 ms. There is, however, a marked difference, because the disks-and-spring system oscillates forever if $\gamma = 0$, while volume oscillations appear to die out slowly for the membrane model, even in the absence of dissipation. This is due to the fact that the disks-and-spring model has a single vibration mode, while the membrane model has a very large number of coupled modes. While the energy is initially deposited in a single, "breathing" mode, it does not remain localized therein, but transfers instead to all other modes of the membrane.

Such oscillations with a characteristic period of a few tenths of a millisecond are, however, not observed experimentally (see fig. 3). This indicates that the liquid exerts a friction on the membrane, which slows down its natural motion and damps the oscillations. It is not easy to predict theoretically the strength of the friction. We consequently performed additional simulations with $E = 7.2$ MPa and increasing values of the dissipation coefficient $\gamma$, in order to determine for which value of $\gamma$ simulations match experiments. Results of simulations performed with three values of $\gamma$ ranging from $2 \times 10^5$ to $10^6$ s$^{-1}$ are shown in fig. 7 on linear (top plot) and logarithmic (bottom plot) scales. It is seen that experimental results are best reproduced for values of $\gamma$ comprised between $5 \times 10^5$ and $10^6$ s$^{-1}$. Oscillations are indeed damped and the walls of the trap relax with the correct characteristic time. We will come back later to this value of the dissipation coefficient.

**5 - Dynamics of the trap door**

The ability of the door of the trap of *Utricularia* to open completely in about 0.5 ms after excitation of the trigger hairs, to close again after a few milliseconds, and to repeat this cycle tens or hundreds of times during the trap's life, is certainly the key and most impressive feature of this plant. At that point, it should be stressed that the word "door" is misleading, since it suggests the rotation of a more or less rigid panel around hinges, while our high-speed video recordings show that the mechanism of the trap of *Utricularia* is completely different. As illustrated in fig. 8, the inversion of the curvature of the door indeed precedes its opening, and not the opposite as previously assumed [18]. It is only after the inversion of curvature has spread over the whole surface that the door opens and water enters the trap. Demonstration that the door of the trap therefore acts as a flexible valve that buckles under the combined effects of pressure forces and the mechanical stimulation of trigger hairs, and not as a panel articulated on hinges, is probably our major result. We propose in this section a model for such a door/valve and discuss the features that are mandatory for it to work correctly.

5.1 - *Geometry of the door*

Keeping with woodwork terminology, the door consists of three essential parts, namely the frame, the threshold and the panel. Examination of the traps with light-sheet fluorescence microscopy indicates that the panel at rest looks like a portion of a prolate ellipsoid, which is attached to the frame along one of the two limiting ellipses and rests on the threshold (when the door is closed) along the other limiting ellipse. At rest, the surface of the threshold is more or less perpendicular to the edge of the panel. Videos furthermore indicate that the frame and the threshold deform very little during setting and firing of the trap. In the model, we therefore considered that both the frame and the threshold are rigid and fixed.

In order to stick to the dimensions of real traps, the panel of the door was therefore modelled as a quarter of a prolate ellipsoid with major radius $a = 300$ µm, minor radius $b = 240$ µm, and thickness $h = 30$ µm. More precisely, the reference geometry of the panel is described by the following equations

$$\frac{x^2 + y^2}{b^2} + \frac{z^2}{a^2} = 1$$
$$x \geq 0$$
$$y \geq 0. \tag{5.1}$$

The ellipse in the $x = 0$ plane represents the frame and is kept fixed. The ellipse in the $y = 0$ plane represents the free edge of the panel. The mesh we used consists of about $M \approx 1100$ facets and $N \approx 550$ vertices. Note that, if we had used such a fine mesh to describe the trap body, then calculations would have become prohibitively long. This is the essential reason why we separated the simulation of the body from that of the door. The minimum energy geometry, which corresponds to a potential energy $E_{pot} \approx -0.08$ nJ, is only marginally deformed compared to eq. (5.1).

The threshold is modelled as a crescent in the $y = 0$ plane. It has two effects. The principal one is to forbid motion towards negative values of *y* of the portions of the panel that rest on it. This is very simply modelled by cancelling the *y*-component of the global force acting on the portions of the panel that lie on the threshold when this component is negative, which amounts to applying a reaction force normal to the threshold. In real traps, the threshold furthermore exerts a friction on the panel during the sliding phase that occurs just after buckling (see below). We neglected this effect in our model, because it only slightly slows down the overall process without modifying the fundamental mechanism that enables the door to open and close repeatedly. From the practical point of view, the inner border of the threshold was modelled as an ellipse with major radius $a = 300$ μm and minor radius $c = 180$ μm. Negative *y*-component of the global force exerted on vertex *j* were therefore cancelled when the coordinates $(x_j, y_j, z_j)$ of this vertex satisfied the condition

$$\frac{x_j^2}{c^2} + \frac{z_j^2}{a^2} \geq 1$$
$$x_j \geq 0$$
$$y_j = 0. \tag{5.2}$$

The equilibrium geometry (pressure difference $\Delta p = 0$, reduced volume $v_{red} = 1$) of the door model is shown in fig. 9a.

### 5.2 - *Door buckling and opening*

As in sect. 4, we first ran several simulations with a dissipation coefficient $\gamma = 0$ and increasing values of the Young's modulus *E*, in order to check whether the model described above has the correct behavior and to determine which value of *E* leads to a realistic critical pressure for buckling. We therefore decreased the pressure inside the door at "slow" rates ranging from 2 to 10 Pa µs$^{-1}$ and integrated Langevin equations (3.15) with a time step $\Delta t = 0.2$ ns. We obtained that for $E = 2.67$ MPa the door deforms very little up to $\Delta p = 15.6$ kPa, while for larger pressure differences, the panel buckles, slides on the threshold and finally swings wide open. This later point will be illustrated shortly. Note that the Young's modulus of the door membrane is only slightly different from that of the trap membrane ($E = 7.2$ MPa) and lies again in the range of values that are commonly measured for parenchymatous tissues (see for example refs. [26-28]). It might also appear as a surprise that the door deforms very little up to $\Delta p = 15.6$ kPa (see fig. 9b), while the membrane of the trap body deforms continuously when $\Delta p$ increases from 0 to 15 kPa (see fig. 5b). As already mentioned at the beginning of sec. 4.1, this marked difference is actually due to the different geometries of the body and the door. The door is convex everywhere and behaves consequently much like a sphere, which sustains pressure without deforming much up to a critical pressure where it undergoes buckling, that is a very abrupt shape transformation caused by a very small pressure increase. In contrast, the body of the trap is not convex everywhere, but displays instead regions with negative curvature. As illustrated in fig. 5, these regions with negative curvature act as seeds from which deflation propagates like a rolling wave when water is pumped outside the trap.

The free motion ($\gamma = 0$) of the door at pressures slightly larger than the critical one is illustrated in fig. 10, which shows the time evolution of the reduced volume $v_{red}$ of the door. The volumes we calculate are signed quantities, because all facets are oriented and an elementary volume is associated to each of them. The elementary volume is positive (respectively, negative) if the scalar product of the vector relating the origin to the center of mass of the facet with the outward normal to the facet is positive (respectively, negative). The volume therefore changes sign when the door crosses the origin. Examination of fig. 10 shows that for $\gamma = 0$ the inversion time predicted by simulations (slightly less than 0.1 ms) is too small compared to the experimental one (around 0.5 ms). We consequently performed additional simulations with $E = 2.67$ MPa and increasing values of the dissipation coefficient $\gamma$, in order to determine for which value of $\gamma$ simulations match experiments. Results of simulations performed with four values of $\gamma$ ranging from $2 \times 10^4$ to $5 \times 10^5$ s$^{-1}$ are shown in fig. 10. It is seen that experimental results are best reproduced for values of $\gamma$ comprised between $2 \times 10^5$ and $5 \times 10^5$ s$^{-1}$. Note that these values of the dissipation coefficient are of the same order of magnitude as the ones that are best adapted to the description of the dynamics of the trap body ($5 \times 10^5$ to $10^6$ s$^{-1}$, see sect. 4).

The dynamics of the door, obtained from simulations performed with a dissipation coefficient $\gamma = 2 \times 10^5$ s$^{-1}$, is illustrated further in figs. 9b-9h. Fig. 9b shows the geometry of the door for a pressure difference $\Delta p$ slightly larger than 15.6 kPa, just before the onset of buckling. Comparison of figs. 9a and 9b shows that the panel is only slightly deformed with respect to its equilibrium geometry at $\Delta p = 0$. This is, of course, due to the fact that pressure forces are balanced by the reaction of the threshold on the free edge of the panel. Figure 9c shows the first indentation, which appears close to the centre of the panel, at the place where trigger hairs are fixed to the door in real *Utricularia* traps. It is worth mentioning that the fact that the first indentation occurs in the *xy* plane is a consequence of the ellipsoid geometry of the door. When modelling the door as a quarter of a sphere instead of a quarter of an ellipsoid, one indeed observes two symmetrical indentations on the sides of the panel, instead of a single one at the centre. In excellent agreement with high-speed videos (see fig. 8), the inversion of curvature then spreads over the panel in about 0.1 ms, but the door is still closed (fig. 9d). At that point, the surface of the panel, which is flattened against the threshold by pressure forces, is dragged across the threshold. This is certainly the step of the opening sequence that depends most on the precise geometry of the door. As can be checked in fig. 10, it corresponds to a decrease of the speed of evolution of $v_{red}$. The duration of this step can, however, be substantially modified by changing the width of the threshold and/or its inclination with respect to the *xz* plane. The surface of the panel flattened against the threshold by pressure forces is also smaller (and the drag time shorter) if the door is not modelled as a quarter of an ellipsoid, but rather as a smaller portion thereof, like for example a sixth or an eighth of an ellipsoid. For some geometries, the only part of the panel which is ever in contact with the threshold is its free (lower) edge, which simply slides on the threshold. At last, let us recall that in real Utricalaria traps this dragging/sliding motion across the threshold is slowed down by friction forces, which we neglect in our simulations. It is only when the free edge of the panel reaches the inner side of the threshold (fig. 9e) that the door really opens and water enters the trap. Inversion of the door then proceeds freely (figs. 9f-9g) till complete inversion is attained (fig. 9h). Comparison of figs. 8 and 9 shows that the door profiles during opening obtained with the membrane model agree qualitatively with the observed ones.

Complete inversion corresponds to a stable equilibrium in our simulations, because we assumed that the difference between pressure forces exerted on the external and internal sides of the membrane is constant. In real *Utricularia* traps, this pressure difference however decreases as water enters the trap and finally vanishes. When pressures are levelled, the door again closes autonomously in about 2.5 ms. Our high-speed video recordings show that closure of the door proceeds through the same steps as opening, but of course in reverse order. We made no attempt to simulate this last step of the opening/closure door mechanism.

In our simulations, the opening mechanism is fired by increasing slowly $\Delta p$ above the critical pressure for buckling ($\Delta p \approx 15.6$ kPa). In real *Utricularia* traps, the pressure difference $\Delta p$ remains instead almost constant once the trap is set, and the mechanism is fired by potential preys touching the trigger hairs. The question whether triggering is purely mechanical (trigger hairs act as levers) or whether it involves a chemical transmission (sensitivity) is still debated [10,11]. In both cases, what physically happens upon triggering can however be visualized by plotting the energy landscape of the system. The upper curve in fig. 11 (labelled $E_{react}$) represents the potential energy of the door along the reaction pathway that leads from the closed to the open position, $1-v_{red}$ being used as the reaction coordinate. This curve was obtained by assuming that the door is initially at rest ($\Delta p = 0$, $v_{red} = 1$, see fig. 9a) and in recording the elastic energy $E_{pot}$ of the system as $\Delta p$ is increased slowly. Deformation of the door is quasi-static up to $1-v_{red} \approx 0.072$ and $\Delta p \approx 15.6$ kPa, so that $E_{react}$ depends very little on the precise value of $\gamma$ up to $1-v_{red} \approx 0.072$. At this value of $1-v_{red}$, buckling occurs and part of the elastic energy of the system is converted into kinetic energy or dissipated at rates that depend on $\gamma$. Therefore, the curve for $E_{react}$ depends more markedly on $\gamma$ for $1-v_{red} > 0.072$ (the curves shown in fig. 11 were obtained with $\gamma = 2\times 10^5$ s$^{-1}$). In this discussion, we are anyway essentially interested in the region $1-v_{red} < 0.072$.

Let us assume that there exists a constant pressure difference $\Delta p$ between the liquids outside and inside the membrane. The energy of the system along the reaction pathway is then $E_{\text{react}} - \Delta p\, v_0 (1 - v_{\text{red}})$, where $v_0$ is the volume of the door at equilibrium at $\Delta p = 0$. This energy is plotted in fig. 11 for five different values of $\Delta p$ ranging from 5 to 25 kPa. As long as $\Delta p$ remains smaller than the critical pressure of 15.6 kPa, the energy landscape actually consists of a minimum located between $1 - v_{\text{red}} = 0$ and $1 - v_{\text{red}} \approx 0.072$ and a barrier located at $1 - v_{\text{red}} \approx 0.072$. Even if the energy of the buckled door is smaller than that of the unbuckled one, the door cannot buckle, because of the barrier. It remains trapped in the well located below $1 - v_{\text{red}} \approx 0.072$ and deforms only slightly, as illustrated in fig. 9b. In contrast, the barrier no longer exists if $\Delta p > 15.6$ kPa, so that the door buckles freely, as observed in our simulations.

In real *Utricularia* traps, the pressure difference $\Delta p$ in set-conditions is probably only very slightly smaller than the critical pressure for buckling. This implies that the barrier hindering buckling along the reaction pathway is very small, too. In this case, the torsion exerted on the membrane when trigger hairs are touched by a potential prey may be sufficient to give the system that tiny amount of extra energy it needs to overcome the barrier and buckle. On the contrary, if chemical transmission (sensitivity) is involved instead of mechanical action [10,11], then the local bending and stretching energy constants of the membrane are temporarily reduced when the trigger hairs are touched. This has the effect of lowering the barrier and letting the door buckle.

### 6 - Conclusion

The underwater traps of *Utricularia* carnivorous plants catch their preys through the repetition of an "active slow deflation / passive fast suction" sequence. In this paper, we presented experimental results and theoretical models aimed at understanding this mechanism. We first showed that a very simple disks-and-spring model enables to extract important information from the experimental results, like the maximum pumping rate, the characteristic pumping time, the hydraulic permeability of the membrane, the average elastic energy stored in the membrane and the maximum velocity of the fluid during the suction phase. We then proposed a more elaborate model that describes the second step of this sequence, that is the ultra-fast suction phase. This model consists of a thin membrane with strain and curvature energy. The only free parameter in the expression of the elastic energy, the Young's modulus *E* of the membrane, is adjusted by requiring that the pressure difference between the outside and the inside of the traps is close to measured values (10-20 kPa) in set-conditions. Obtained values of *E* (2 to 10 MPa) lie in the range of values that are commonly measured for parenchymatous tissues. The door of the trap is modelled as a quarter of an ellipsoid, one edge of which is fixed, while the other one is free and rests on the threshold in set-conditions. Our simulations show that, for a pressure difference slightly larger than the critical one, the door buckles, slides on the threshold and finally swings wide open. This sequence is in excellent agreement with that observed in high-speed videos (fig. 8).

This model therefore strongly supports the hypothesis that we formulated by looking at the high-speed videos, that is, that the trap acts as a flexible valve that buckles under the combined effects of pressure forces and the mechanical stimulation of trigger hairs, and not as a panel articulated on hinges. The only real limitation of this model is that the liquid is only roughly taken into account through the dissipation coefficient $\gamma$ in Langevin equations. It was shown that $\gamma$ must be chosen in the range $2\times10^5$ to $10^6$ s$^{-1}$ in order for the characteristic times of the model to match observed ones. A better model would consist in taking water explicitly into account and in integrating coupled equations for the dynamics of the liquid and the membrane. This is, however, a much more complex problem.

To conclude, let us note that this work on the underwater ultra-fast traps of *Utricularia* opens very interesting perspectives for the practical design of flexible structures performing fast motion in a fluid. Since such flexible structures show less fatigue than articulated ones, the mechanism of the tiny traps of *Utricularia* suggests a new kind of microfluidic tools, based on buckling, for Lab-on-chip devices.

# REFERENCES

[1] D. Attenborough, *The private life of plants* (BBC Books, 1994) (ISBN 0-563-37023-8) and BBC documentary, January 1995.

[2] L. Gaume, Y. Forterre, PLoS ONE 2(11), e1185 (2007)

[3] Y. Forterre, J. M. Skotheim, J. Dumais, L. Mahadevan, Nature 433, 421 (2005)

[4] B. E. Juniper, R. J. Robins, D. M. Joel, *The Carnivorous Plants* (Academic, London, 1989)

[5] F. E. Lloyd, *The Carnivorous Plants* (Waltham, Massachusetts, 1942)

[6] A. T. Czaja, Zeitschr. f. Bot. 14, 705 (1922)

[7] T. Ekambaram, Journ. Indian. Bot. Soc. 4, 73 (1924)

[8] F. E. Lloyd, Can. J. Bot. 10, 780 (1932)

[9] F. E. Lloyd, Plant Physiol. 4, 87 (1929)

[10] T. Diannelidis, K. Umrath, Protoplasma 42, 58 (1953)

[11] P. Sydenham, G. Findlay, Australian J. Biol. Sci. 26, 1115 (1973)

[12] C. L. Withycombe, Jour. Linn. Soc. 46, 401 (1924)

[13] A. Sasago, T. Sibaoka, Bot. Mag. Tokyo 98, 55 (1985)

[14] A. Sasago, T. Sibaoka, Bot. Mag. Tokyo 98, 113 (1985)

[15] P. Taylor, *The genus Utricularia: a taxonomic monograph*. (Kew Bulletin Additional Series XIV, London, 1989)

[16] K. Reifenrath, I. Theisen, J. Schnitzler, S. Porembski, W. Barthlott, Flora 201, 597 (2006)

[17] O. Vincent, C. Weisskopf, S. Poppinga, T. Masselter, T. Speck, M. Joyeux, C. Quilliet, P. Marmottant, submitted to Proceedings of the Royal Society B : Biological Sciences

[18] J.M. Skotheim, L. Mahadevan, Science 308, 1308 (2005)

[19] *Introduction to Particle Technology*, edited by M. Rhodes (Wiley, Weinheim, 1998)

[20] F.I. Niordson, *Shell theory* (North Holland, New-York, 1985)

[21] S. Komura, K. Tamura, T. Kato, Eur. Phys. J. E 18, 343 (2005)

[22] A.F. Bower, *Applied mechanics of solids* (CRC Press, 2009) (ISBN 978-1439802472)

[23] M. Meyer, M. Desbrun, P. Schröder, A.H. Barr, *Discrete differential geometry operators for triangulated 2-manifolds*, in *Visualization and mathematics III*, edited by H.C. Hege and K. Polthier (Springer, Heidelberg, 2003), p. 35-57

[24] J. Yong, B. Deng, F. Cheng, B. Wang, K. Wu, H. Gu, Sci. China Ser. F - Inf. Sci. 52, 418 (2009)

[25] E. Magid, O. Soldea, E. Rivlin, Computer Vision and Image Understanding 107, 139 (2007)

[26] H. Alizadeh, L.H. Segerlind, Applied Engineering in Agriculture 15, 507 (1997)

[27] L. Mayor, R.L. Cunha, A.M. Sereno, Food Research International 40, 448 (2007)

[28] K.J. Niklas, Amer. J. Bot. 75, 1286 (1988)

**FIGURE CAPTION**

**Figure 1** : (a) Stereo microscopy view of an *Utricularia inflata* trap. The door and the trigger hairs face the right edge of the picture. The other two pictures show lateral views of the door in closed (b) and open (c) positions, which were obtained with an ultrafast camera. The black shadow at the upper right edge of the pictures is the lever, which is used to manually excite trigger hairs and fire the trap mechanism.

**Figure 2** : (Color online) Deflation of the trap body. The plots show the evolution of the thickness $e$ of the trap (expressed in mm) as a function of time (expressed in minutes) on linear (top plot) and logarithmic (bottom plot) scales. The trap is fired manually at time $t = 0$ and measurement of $e$ starts immediately after the ultra-fast closing of the door. The insert in the top plot shows a trap close to maximum deflation viewed from above and indicates where the thickness $e$ is measured. The door of the trap faces the right edge of the figure. The dot-dashed line in the bottom plot shows the result of the least square adjustment with $\tau_{pump} = 53$ minutes.

**Figure 3** : (Color online) Inflation of the trap body after triggering. The plots show the evolution of the thickness $e$ of the trap (expressed in mm) as a function of time (expressed in ms) on linear (top plot) and logarithmic (bottom plot) scales. The origin of the time scale is somewhat arbitrary. The dot-dashed line in the bottom plot shows the evolution of an exponential process with time constant $\tau = 1.3$ ms.

**Figure 4** : Stereo microscopy views of two cuts of the *Utricularia inflata* trap. (a) Transverse section of the trap body. The sharp kink, which is observed in the right part of the figure, is due to the fact that the membrane was slightly damaged during the cut. (b) View of the door (separated from the rest of the trap) seen from below the edge that rests on the threshold.

**Figure 5** : (Color online) Simulated trap body (without the door). The left figure ($V_{red} = 1.0$) represents the minimum energy geometry, that is, the equilibrium geometry of the trap when $\Delta p = 0$. The right figure ($V_{red} = 0.6$) corresponds to the trap in set-conditions, when it is ready to fire. This is the equilibrium geometry for $\Delta p \approx 15$ kPa, and the initial condition for the inflation simulations reported in sect. 4.3.

**Figure 6** : (Color online) Simulation of the deflation of the trap body. The top and bottom plot show the evolution of $\Delta p$ and $E_{pot}$, respectively, as a function of $1 - V_{red}$. The pressure difference between the outside and the inside of the trap, $\Delta p$, is expressed in kPa, and the elastic energy stored in the membrane, $E_{pot}$, in µJ. As indicated on the plots, the various curves were obtained either by integrating Langevin equations with $\gamma = 0$ (blue dot-dash line) and $\gamma = 10^4$ s$^{-1}$ (red solid line) or by minimizing $E_{pot}$ for each value of $V_{red}$ (green short-dash line). The brown long-dash line in the bottom plot was also obtained by integrating Langevin equations with $\gamma = 10^4$ s$^{-1}$, but for a mesh with only about 200 vertices instead of 2150 ones. For dynamics simulations, the pressure inside the trap was decreased at the rate of 1 Pa µs$^{-1}$, while Langevin equations were integrated numerically with a time step $\Delta t = 10$ ns.

**Figure 7** : (Color online) Simulation of the inflation of the trap body. The plots show the evolution of the reduced volume $V_{red}$ of the trap as a function of time (expressed in ms) on linear (top plot) and logarithmic (bottom plot) scales. The trap is assumed to be initially at equilibrium with the geometry shown in the right picture of fig. 5 ($V_{red} = 0.6$, $\Delta p \approx 15$ kPa). $\Delta p$ is instantly switched to 0 at time $t = 0$ and Langevin equations are integrated numerically with a time step $\Delta t = 2.5$ ns. As indicated on the plots, the various curves were obtained with four different values of $\gamma$ ranging from 0 to $10^6$ s$^{-1}$. The dot-dashed line in the bottom plot shows the time evolution of an exponential process with a characteristic time $\tau = 1.3$ ms, for the sake of an easier comparison with the bottom plot of fig. 3.

**Figure 8** : High-speed recording of the door opening of an *Utricularia australis* trap after manual triggering with a needle. These fluorescence images were captured at 2900 frames per second using a microscope equipped with a laser sheet illumination apparatus, which enables to image only a thin slice of the living trap (see ref. [17] for more information). The figure displays a selection of images at 0, 5.9, 7.6, 7.9, 8.6 and 9.7 ms after the first door motion. (a) shows the trap in set conditions. The inversion of curvature spreads gradually on the whole door ((b) to (d)) before the door opens wide ((e)) and closes back ((f)). The speed of aperture of this *Utricularia australis* trap is significantly slower than that of the *Utricularia inflata* traps.

**Figure 9** : (Color online) Snapshots of the opening dynamics of the simulated trap door. (a) shows the geometry of the door at equilibrium ($\Delta p = 0$, $v_{red} = 1$). (b) to (h) show the opening of the door when submitted to a pressure difference slightly larger than the critical pressure for buckling ($\Delta p \approx 15.6$ kPa). The origin of times, $t = 0$, is somewhat arbitrary. Langevin equations (3.15) were integrated numerically with a time step $\Delta t = 0.2$ ns and a dissipation coefficient $\gamma = 2 \times 10^5$ s$^{-1}$.

**Figure 10** : (Color online) Simulation of the opening of the trap door. The plot shows the time evolution of the reduced volume $v_{red}$ of the door when it is submitted to a pressure difference slightly larger than the critical pressure for buckling ($\Delta p \approx 15.6$ kPa). The origin of times, $t = 0$, is somewhat arbitrary. Langevin equations (3.15) were integrated numerically with a time step $\Delta t = 0.2$ ns. As indicated on the plots, the various curves were obtained with five different values of $\gamma$ ranging from 0 to $5 \times 10^5$ s$^{-1}$. For $\gamma = 5 \times 10^5$ s$^{-1}$, complete inversion ($v_{red} \approx -1$) is achieved in about 1 ms, as shown in the small insert.

**Figure 11** : (Color online) Energy landscape of the trap door along the reaction pathway for opening. $1 - v_{red}$ is used as the reaction coordinate. The upper curve, labelled $E_{react}$, shows the elastic energy of the door along the reaction pathway for door opening. The five other curves show the actual energy of the system along the reaction pathway, $E_{react} - \Delta p\, v_0 (1 - v_{red})$, for five different values of $\Delta p$ ranging from 5 to 25 kPa. Buckling is forbidden for values of $\Delta p$ smaller than 15.6 kPa by the energy barrier at $1 - v_{red} \approx 0.072$.

**FIGURE 1**

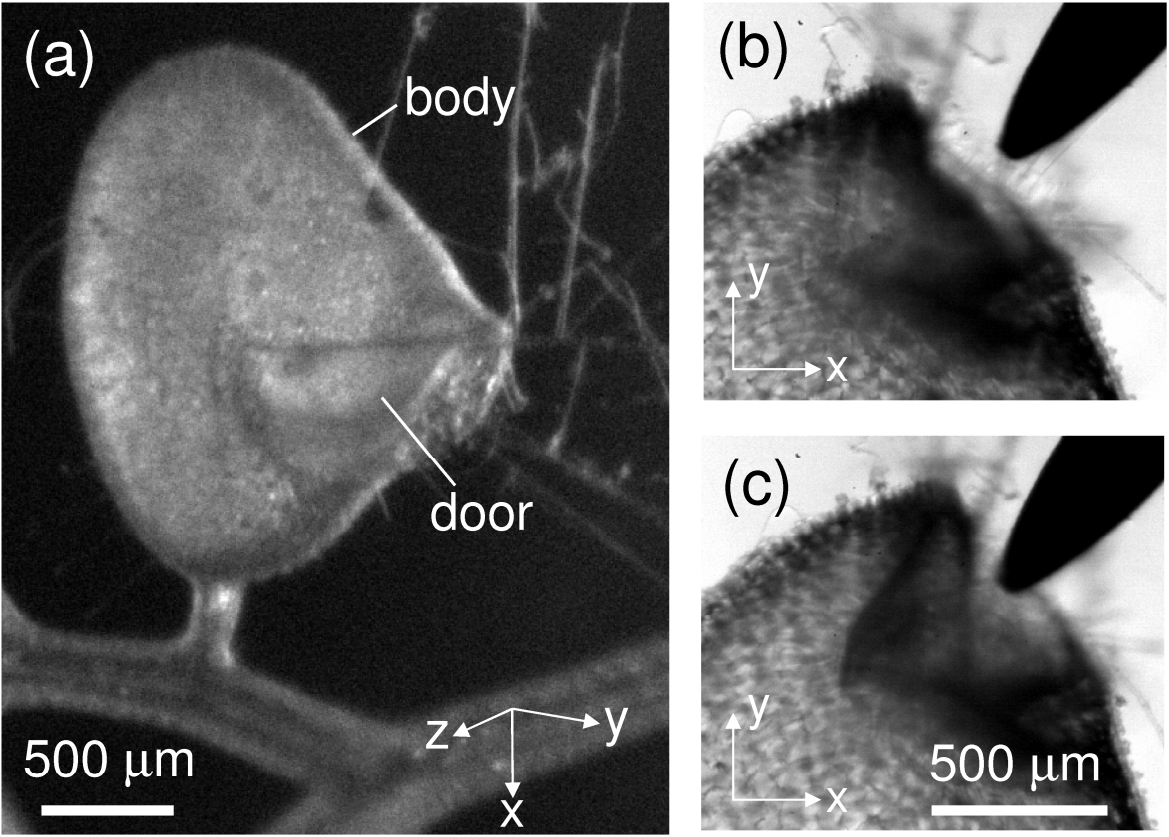

**FIGURE 2**

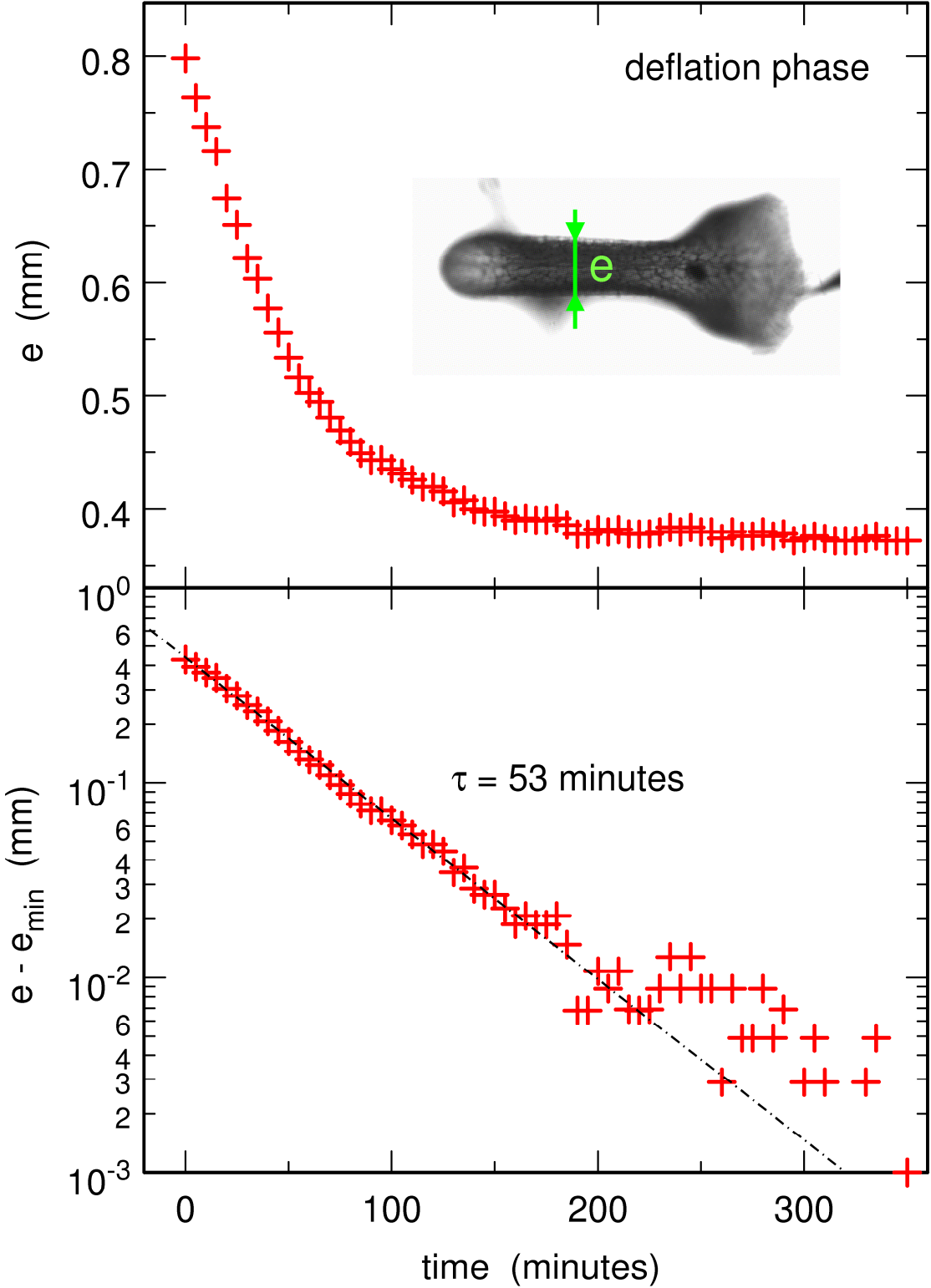

**FIGURE 3**

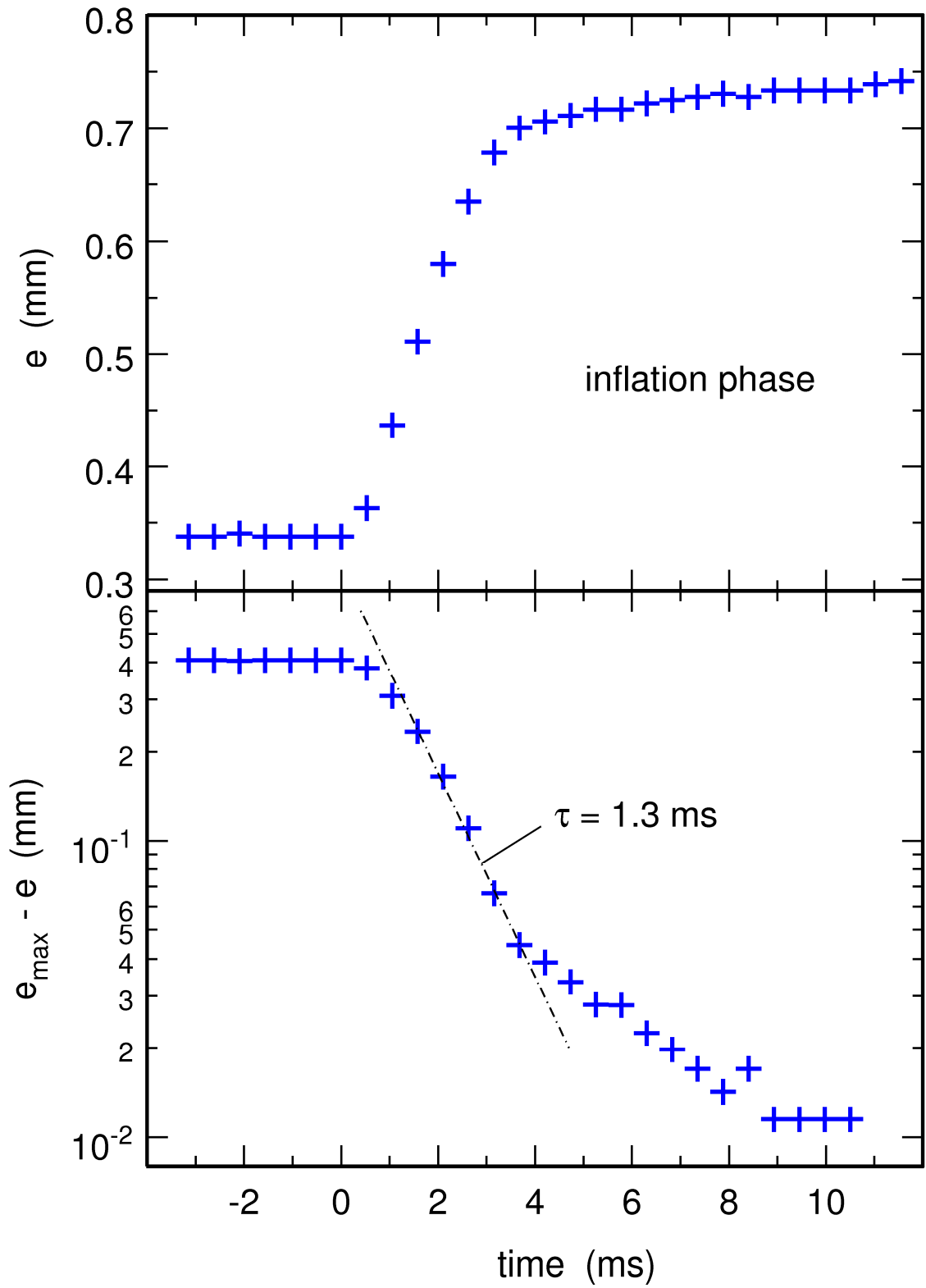

**FIGURE 4**

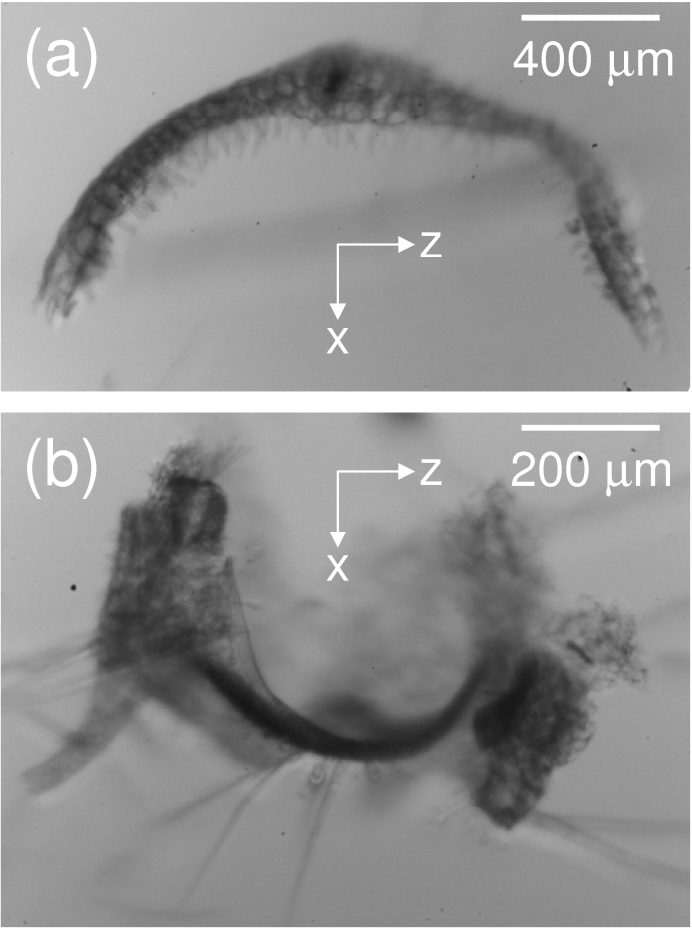

**FIGURE 5**

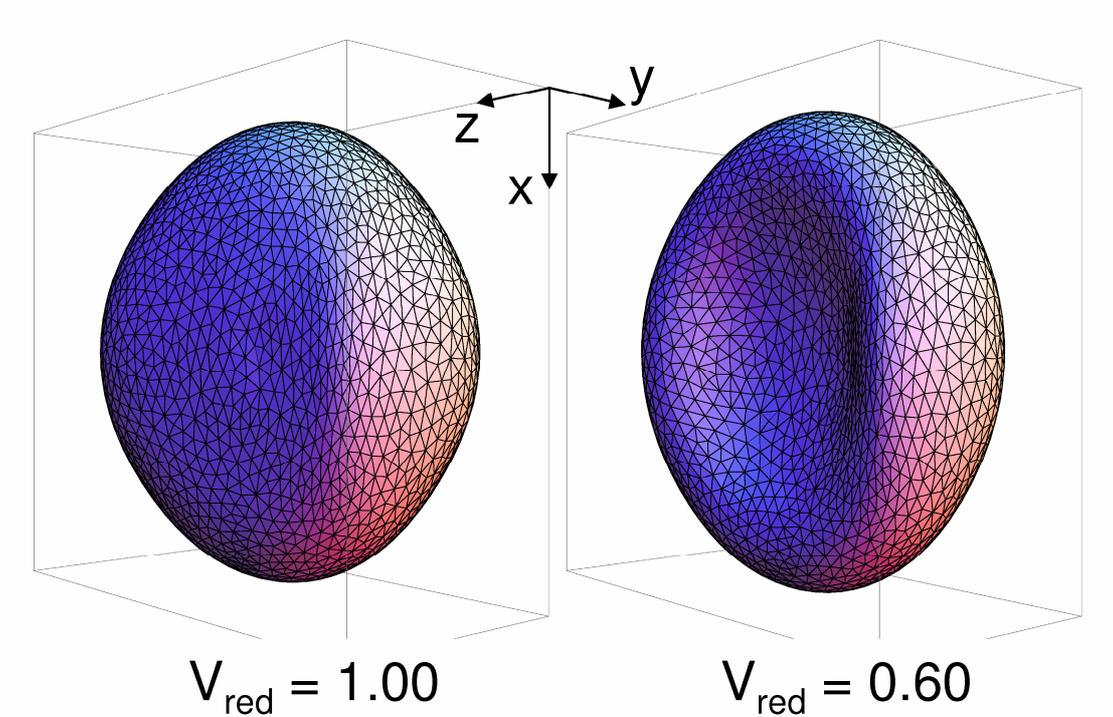

$V_{red} = 1.00$   $V_{red} = 0.60$

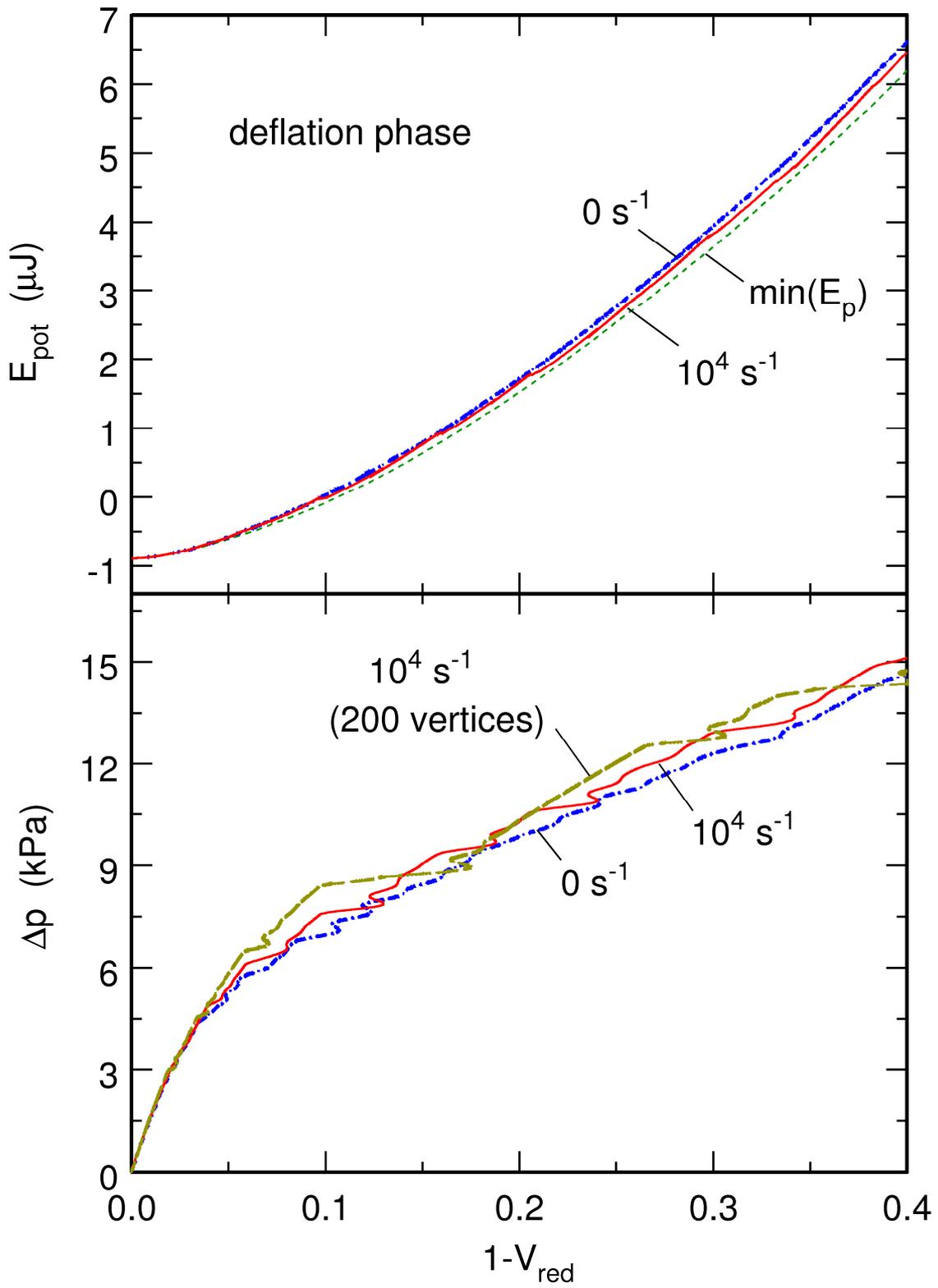

**FIGURE 7**

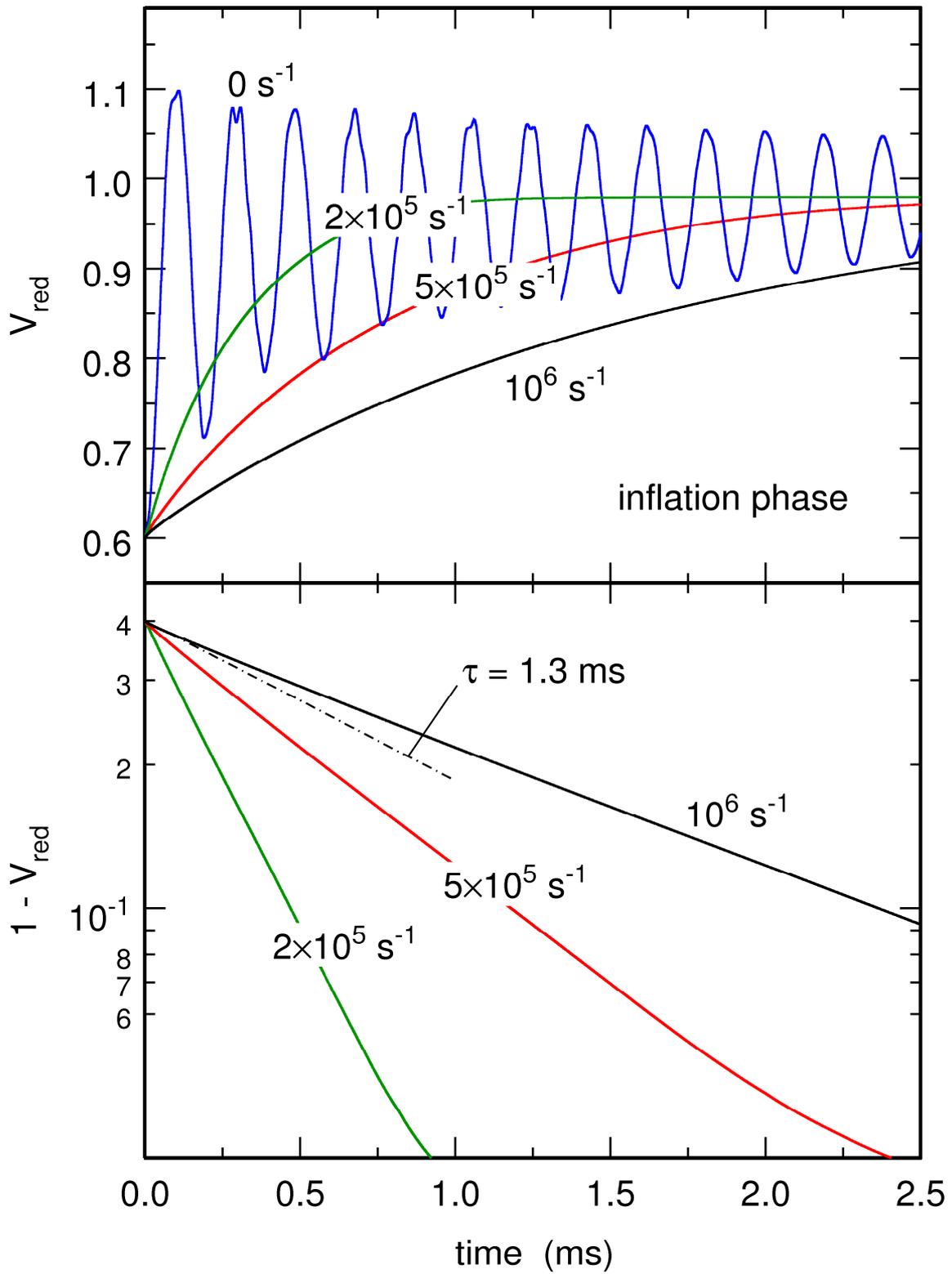

**FIGURE 8**

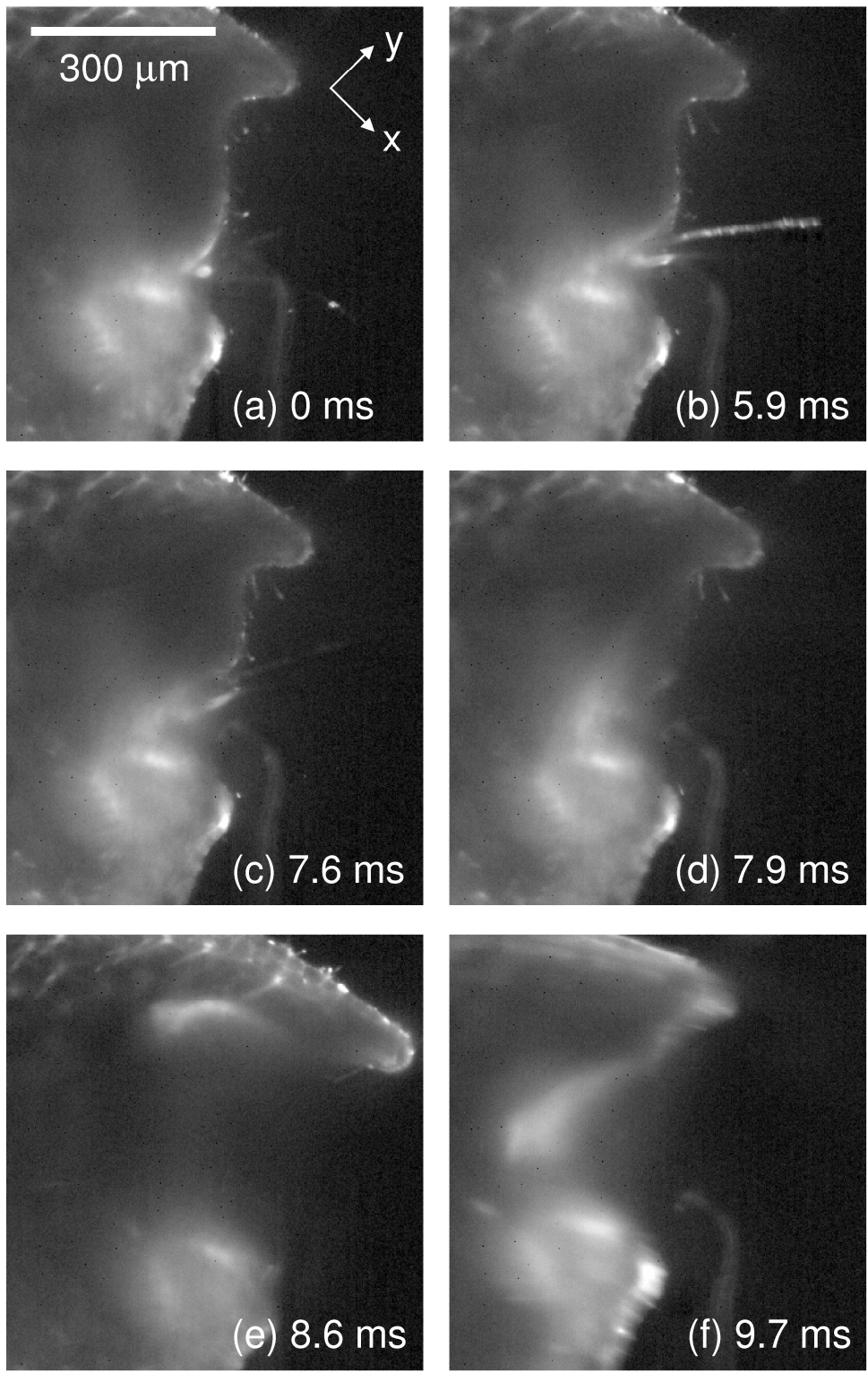

FIGURE 9a) ($\Delta p = 0$); $v_{red} = 1.00$
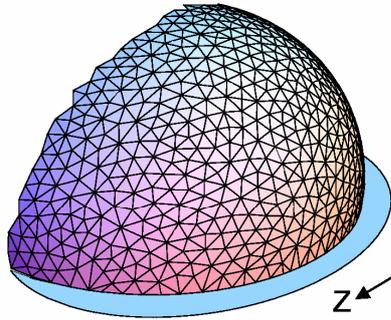

b) $t = 15$ µs; $v_{red} = 0.93$
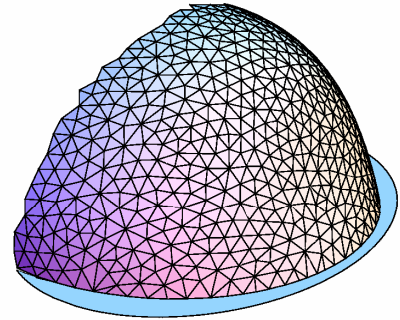

c) $t = 30$ µs; $v_{red} = 0.92$
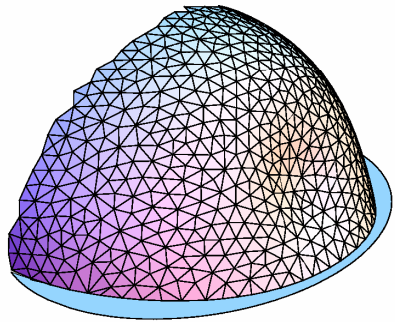

d) $t = 120$ µs; $v_{red} = 0.44$
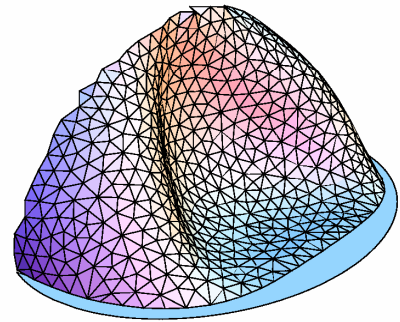

e) $t = 200$ µs; $v_{red} = 0.23$
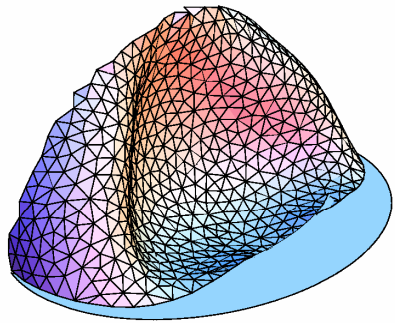

f) $t = 300$ µs; $v_{red} = -0.08$
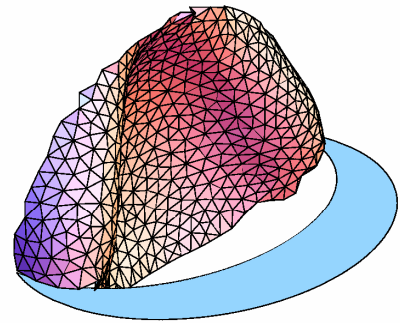

g) $t = 360$ µs; $v_{red} = -0.57$
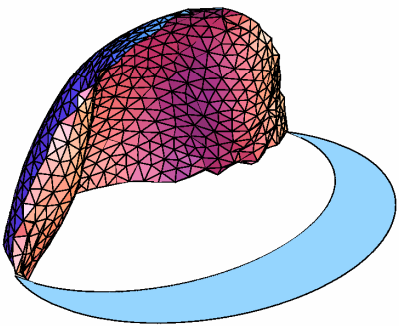

h) $t = 420$ µs; $v_{red} = -1.02$
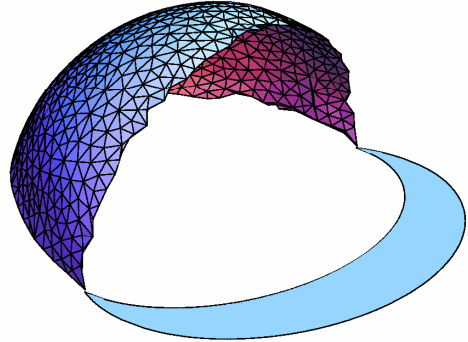

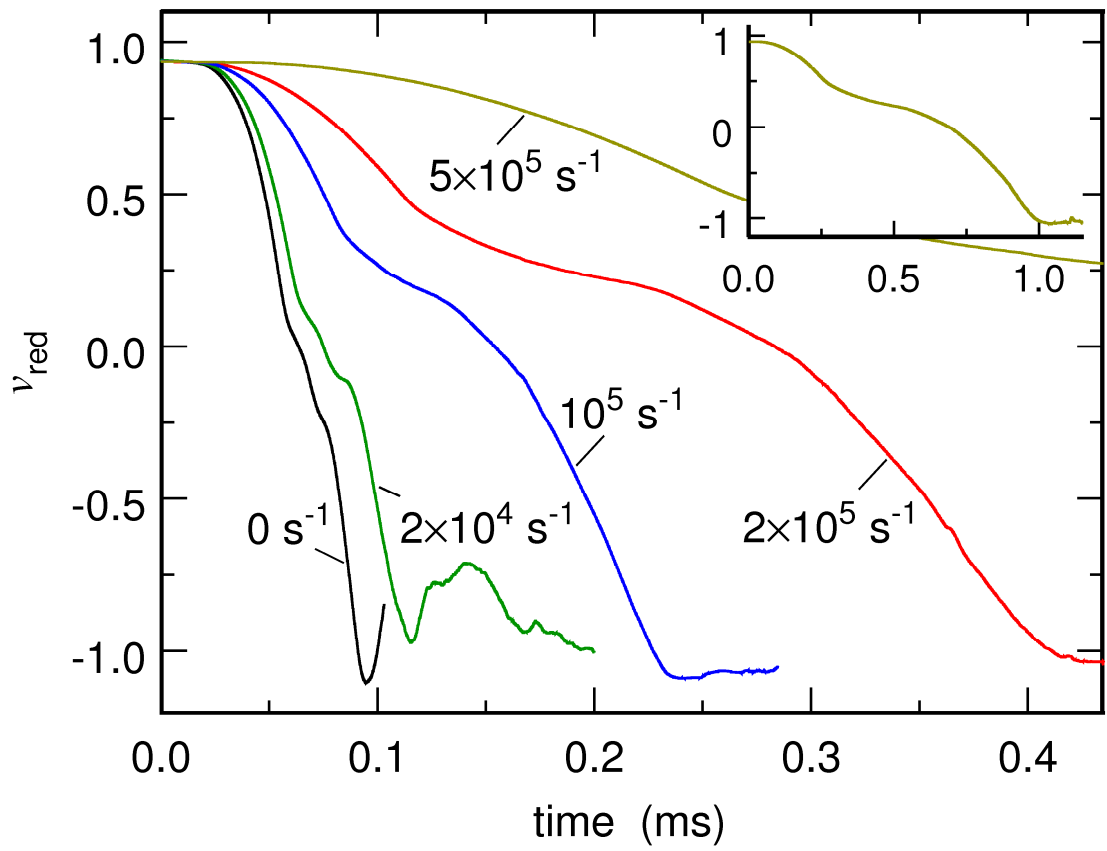

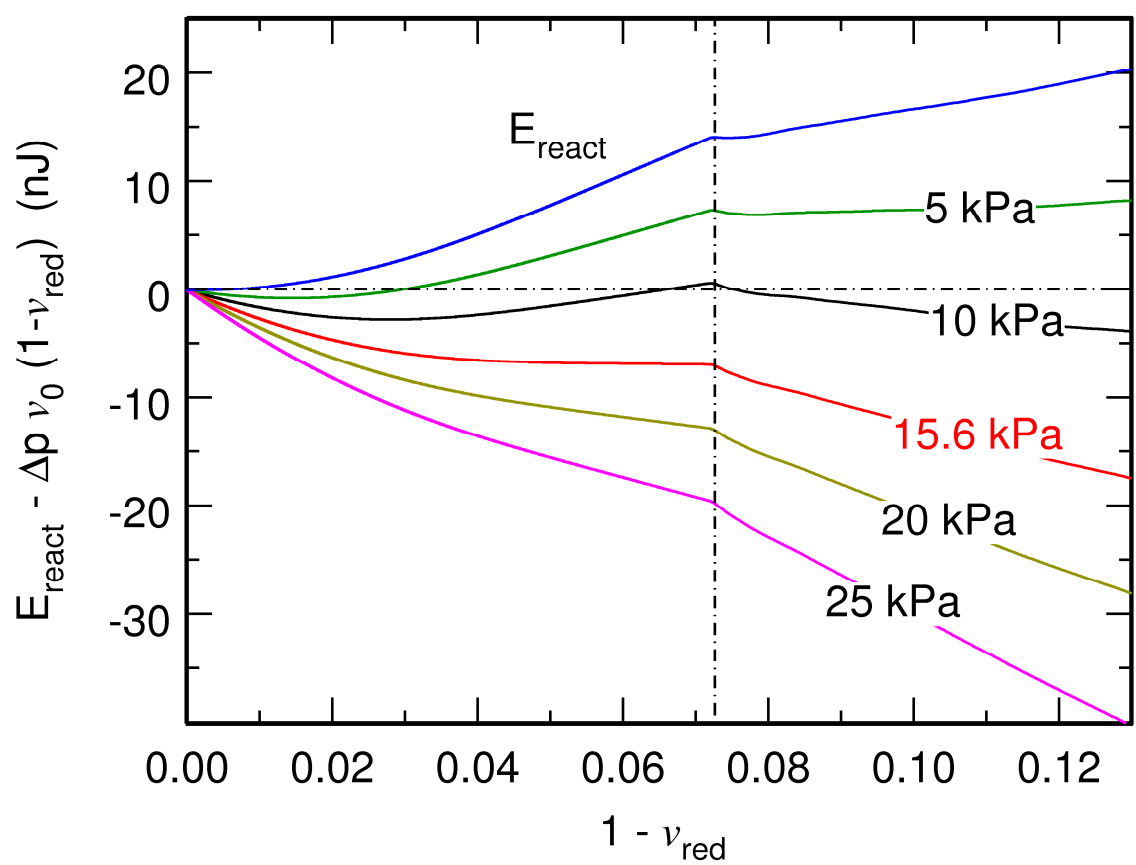